\documentstyle[sprocl,epsfig]{article}

\newcommand{\gsim}{\raisebox{-0.07cm}{$\:\stackrel{>}{{\scriptstyle
 \sim}}\: $} }
\newcommand{\lsim}{\raisebox{-0.07cm}{$\:\stackrel{<}{{\scriptstyle
 \sim}}\: $} }
\newcommand{\als}{\alpha_s}
\newcommand{\ab}{\overline{\alpha}_s}
\newcommand{\ra}{\rightarrow}
\begin{document}

\small
\vspace*{-0.5cm}
\begin{flushleft}
WUE-ITP-98-031  \hfill {\tt hep-ph/9807369} \\
July 1998
\end{flushleft}
\normalsize
\vspace*{0.6cm}

\title{DIS$\,$'98 STRUCTURE FUNCTIONS SUMMARY, PART II\\[-2mm]} 

\author{ANDREAS VOGT \\[1mm]}

\address{Institut f\"ur Theoretische Physik, Universit\"at W\"urzburg,
\\ Am Hubland, D-97074 W\"urzburg, Germany 
\\E-mail: avogt@physik.uni-wuerzburg.de\\[1mm]}


\maketitle\abstracts{Recent results presented in the structure 
functions working group are briefly summarized for the following 
topics: The theoretical treatment of heavy quarks in structure 
functions,  higher-order corrections for the leading-twist evolution 
(inclu\-ding small-$x$ resummations), the present status of the 
proton's parton densities, and the impact of higher twists on 
determinations of the strong coupling constant. The reader is referred 
to Part I~\cite{WG1a} for accounts of the transition to the 
photoproduction region, pion and photon structure results, and
high-$Q^2$ phenomena.\\[-4mm]}

%%%%%%%%%%%%%%%%%%%%%%%%%%%%%%%%%%%%%%%%%%%%%%%%%%%%%%%%%%%%%%%%%%%%%%%
\section{Heavy quarks in structure functions}
%%%%%%%%%%%%%%%%%%%%%%%%%%%%%%%%%%%%%%%%%%%%%%%%%%%%%%%%%%%%%%%%%%%%%%%

The charm structure functions, especially $F_2^c$, have attracted 
considerable interest over the past years. Unlike in the fixed-target 
regime, $F_2^c$ makes up a sizeable fraction, up to about a quarter, 
of the total $F_2$ in the HERA \mbox{small-$x$} region. Despite being 
suppressed, it contributes significantly as well to the scaling 
violations in the kinematic range covered by the NMC data. 

\vspace{1mm}
At low scales, $Q^2 \approx m_c^2$, $F_2^c$ is uniquely calculated from 
the light parton densities via the $\gamma^{\ast}g \ra c\bar{c}$ 
Bethe-Heitler process and its $O(\als^2)$ corrections~\cite{LSRN}, 
without invoking the concept of a charm parton distribution (we neglect 
here a possible intrinsic charm component, which seems to be relevant 
only at high~$x$). For $Q^2 \gg m_c^2$ large logarithms 
appear in the coefficient functions $C_{2,L}$, which may require a
resummation. At $x < 10^{-2}$ these logarithms dominate $C_2$ already 
for $Q^2 > 20 $ GeV$^2$, but $C_L$ only above 10$^{3}\,$GeV$^2$~\cite
{ASYM}. Previous leading-order results~\cite{ACOT} for this resummation 
have been extended to higher orders in ref.~\cite {F4N}. This leads to 
a high-$Q^2$ description in terms of four massless flavours, with the 
charm distribution uniquely specified by the light parton densities. 

\vspace{1mm}
The problem of the transition between both approaches, i.e., the 
construction of a variable flavour-number scheme (VFNS) has also been
addressed~\cite{F4N,RT}. There seems to be agreement that a unique 
construction does not exist, and indeed the prescription of refs.\ 
\cite{F4N,RT} differ at non-asymptotic values of $Q^2$. Note, however, 
that this ambiguity concerns only the coefficient functions, as the 
parton evolution can be kept strictly massless without any loss of
generality~\cite{OS}. Results for the VFNS prescription of ref.~\cite
{F4N} are compared to the non-resummed calculation (usually 
called fixed flavour-number scheme, FFNS) in Fig.~1. The differences 
are typically 10\% at small $x$, i.e., they are of the same size as the 
factorization-scale dependence of the FFNS calculations \cite{AVc}, 
shown for one value of $Q^2$ in Fig.~1 as well. Hence also the latter 
approach seems to be applicable at the present level of accuracy in the 
HERA small-$x$ region. On the other hand, somewhat larger effects are 
possible for other VFNS prescriptions~\cite{RT}.
\begin{figure}[h]
\begin{minipage}[b]{11.9cm}
\vspace*{-1mm}
~\parbox[c]{5.8cm}{
\centerline{\epsfig{file=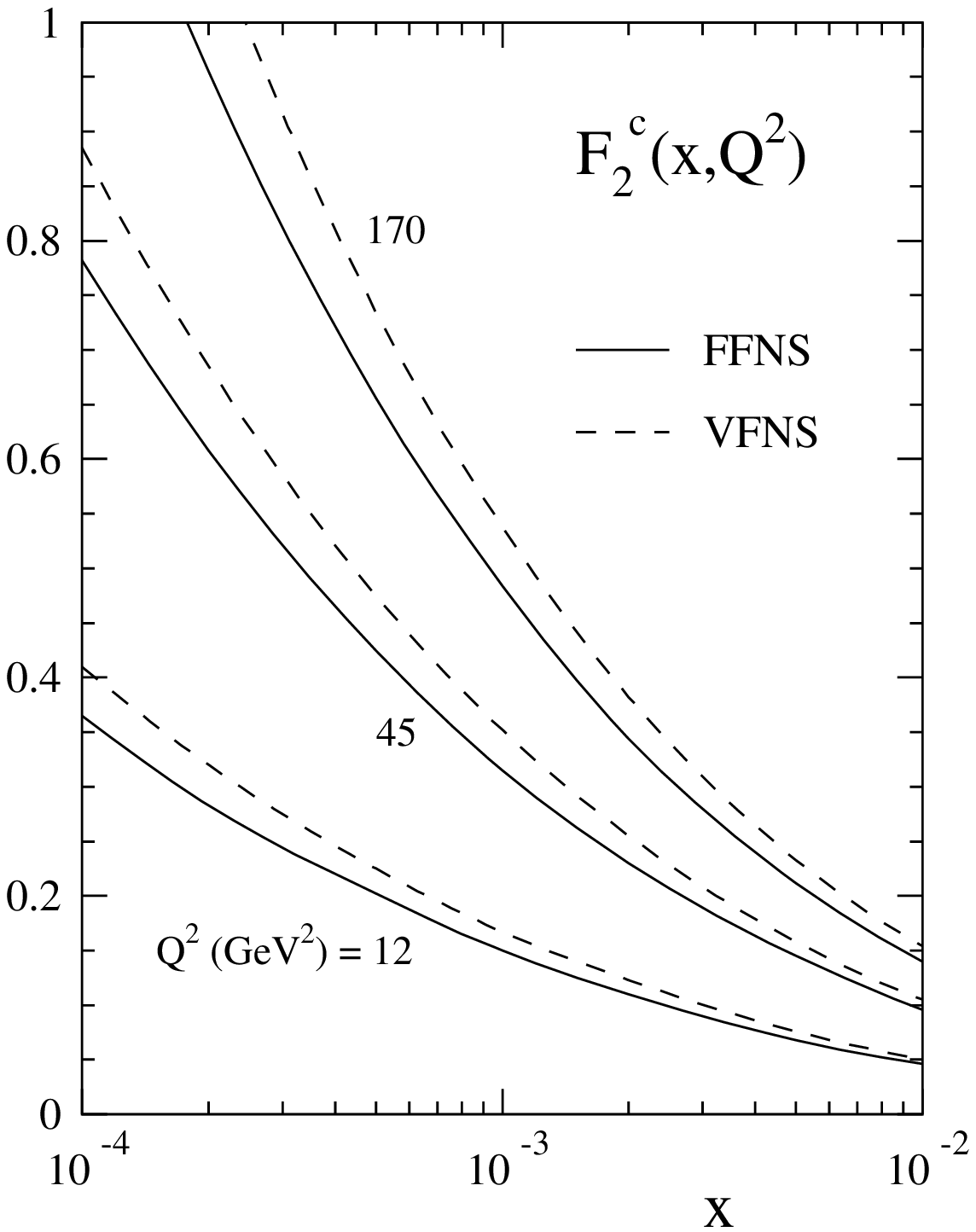,width=5.8cm,angle=0}}
}~
\parbox[c]{5.8cm}{
\centerline{\epsfig{file=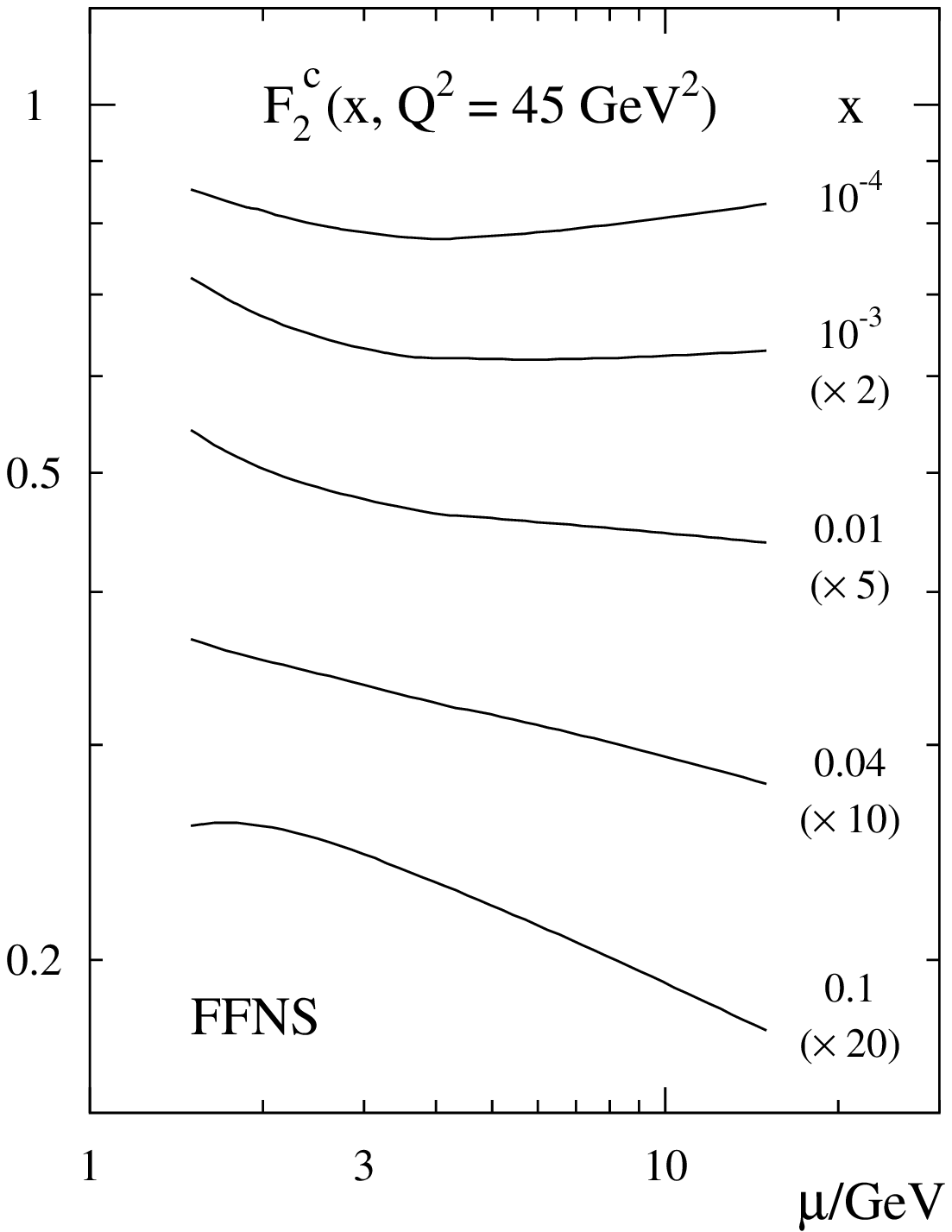,width=5.8cm,angle=0}}
}
\\ \mbox{}
\end{minipage}

\noindent
{\small\sf Figure~1: Left part: comparison of $F_2^c$ as obtained in 
NLO from FFNS and VFNS calculations~\cite{F4N}. Right part: the 
factorization scale dependence of the FFNS predictions at one 
representative $Q^2$.  The light parton densities in both parts are not 
the same.}
\end{figure}

Until now the extraction of $F_2^c$ from $D^{\ast}$ production at HERA 
has been performed only in the non-resummed approach (FFNS), as the 
required exclusive cross sections have been calculated only in this 
framework so far~\cite{HS}. However, first results have been 
presented from a VFNS event-generator Monte-Carlo program for 
semi-inclusive heavy-quark production in DIS~\cite{Ol1}.

\vspace{1mm}
As is obvious from Fig.~1, the stability of the calculations 
deteriorates towards larger $x$, i.e., towards the threshold region. 
This effect is even more pronounced at lower $Q^2$, and renders present 
calculations in the range of fixed-target experiments rather 
unreliable. A Sudakov resummation of leading and next-to-leading 
threshold logarithms, as discussed in ref.~\cite{Moch}, may lead to a 
stable framework also in this regime. 

%%%%%%%%%%%%%%%%%%%%%%%%%%%%%%%%%%%%%%%%%%%%%%%%%%%%%%%%%%%%%%%%%%%%%%%
\section{Higher-order corrections for structure function evolution}
%%%%%%%%%%%%%%%%%%%%%%%%%%%%%%%%%%%%%%%%%%%%%%%%%%%%%%%%%%%%%%%%%%%%%%%

We confine ourselves to the unpolarized singlet case, where new results 
have been obtained during the last year. The corresponding 2-loop (NLO) 
splitting functions are fully (at all~$x$) known for about two decades. 
On the other hand, at $O(\als^3)$ only the four lowest even-integer 
moments were determined so far \cite{LNRV}. A first step towards the 
full 3-loop anomalous dimensions, which are required to match the 
accuracy of present and forthcoming high-precision data, has been taken 
recently by calculating the finite terms of the 2-loop operator matrix 
elements~\cite{MSN}. Moreover a first partial result has been derived 
by means of the large-$N_f$ expansion, namely the $(\als /4\pi)^3 N_f^2 
C_G$ contribution to $P_{gg}$ \cite{BG}:
\begin{eqnarray*}
 \lefteqn{P^{\,\footnotesize{\mbox{$3$-loop}}}_{gg}(x,N_f^2 C_A)~ =~ - 
 \frac{1}{54} \Big[ 87\,\delta(1-x) ~+~ ( 304 + 172x + 208x^2)\ln x} \\
& & -~ 48(1+x) \ln^2 x ~+~ 32 ~-~ \frac{32}{[1-x]_+} +~ 192(1+x) 
 [\psi^\prime(1) - \mbox{Li}_2(x)] \\
& & +~ \frac{4(1-x)}{x} (52 + 19x + 52x^2) \ln(1-x) +~\frac{4(1-x)}{3x}
 (236 + 47x + 236x^2) \Big] \, .
\end{eqnarray*}
The first moments of this expression agree with the results of ref.\ 
\cite{LNRV}.

\vspace{1mm}
An alternative approach for the small-$x$ region has been to resum the 
most singular small-$x$ terms of $P_{ij}$ to all orders in $\als $. For
the gluonic splitting functions $P_{gq}$ and $P_{gg}$ these terms read
$c_k^{\,\rm Lx} \cdot (1/x) \alpha_s^k \ln (1/x)^{k-1}$, and the 
coefficients $c_k^{\,\rm Lx}$ were determined long ago as well. More 
recently also the leading contributions to $P_{qq}$ and $P_{qg}$ were 
derived \cite{CH}, which contain one power of $\ln (1/x)$ less than 
their gluonic counterparts. These terms dominate the respective 
splitting functions at some very low values of $x$, depending on the 
size of the less singular contributions. Until recently estimates of 
the impact of such terms, which tends to be enhanced substantially by 
the ubiquitous Mellin convolution, were only possible by educated 
guesses based on momentum conservation constraints and the structure of 
the LO and NLO splitting functions~\cite{EHW}.

\vspace*{1mm}
An inclusion of the leading small-$x$ logarithms into the analysis can 
lead to very good fits of all small-$x$ $F_2$ and $F_2^c$ HERA data, as 
demonstrated in a scheme-independent evolution-equation approach 
\cite{Th2} all well as within the framework of $k_{\perp}$ factorization
\cite{Mun}. However, both approaches seem to require values for 
observables and parameters  -- a rather small $F_L$ in the first case 
and, more notably, a very small $\alpha_s$ in the second one -- which 
may be interpreted as phenomenological indications of large subleading 
corrections.

\vspace*{1mm}
During the past year the calculation of $O(\als )$ corrections to the 
BFKL kernel has been completed \cite{FL,CC}. This result fixes the 
next-to-leading small-$x$ (NL$x$) piece of $P_{gg}$. In the DIS scheme
the presently known terms read~\cite{BRNV}
$$
  P_{gg}^{\,\rm DIS}(N,\als ) = \ab P_{gg,0} (N) + \ab^2 P_{gg,1}
  ^{\,\rm DIS}(N) + \sum_{l=3}^{\infty} \left( \frac{\ab}{N}\right)^{l} 
  \Big( b^{\,\rm Lx}_{gg,l} + N\, b^{\,\rm NLx}_{gg,l} \Big) \: ,
$$
with $\ab = 3\als/\pi$ and N the usual Mellin variable shifted by one
unit. The  L$x$ and NL$x$ coefficients $b_{gg,l}$ are compared in
Fig.~2, where also the resulting splitting function is shown for $\als 
= 0.2 $. The NL$x$ corrections turn out to be exceedingly large in the 
HERA $x$-region, leading to a grossly negative splitting function 
already above $x \simeq 10^{-3}$. Thus the $\ln (1/x)$ expansion is
inapplicable to $P_{gg}$ at any $x$-values of practical interest, a 
situation which is by no means a special feature of the DIS scheme as 
demonstrated in ref.~\cite{BF}. 

\vspace*{1mm}
Likewise, taking the NL$x$ corrections at face value, the hard pomeron 
intercept $\omega_P $ would read $ \:\omega_P (\als) = 2.65 \,\als 
(1 - 6.36\,\als )\: $ for $N_f = 4$ \cite{BRNV}, leading to negative 
values for $Q^2$ as high as about 300 GeV$^2$. Hence a reliable 
extension of structure-function evolution calculations beyond the usual 
fixed-order perturbation theory, if possible at all, will require new 
theoretical concepts.
%
%Hence new concepts need 
%to be introduced for reliably extending structure-function evolution
%calculations beyond the usual fixed-order perturbation theory 
%\cite{CC}. 

\begin{figure}
\begin{minipage}[b]{11.9cm}
\parbox[c]{5.6cm}{
\mbox{\epsfig{file=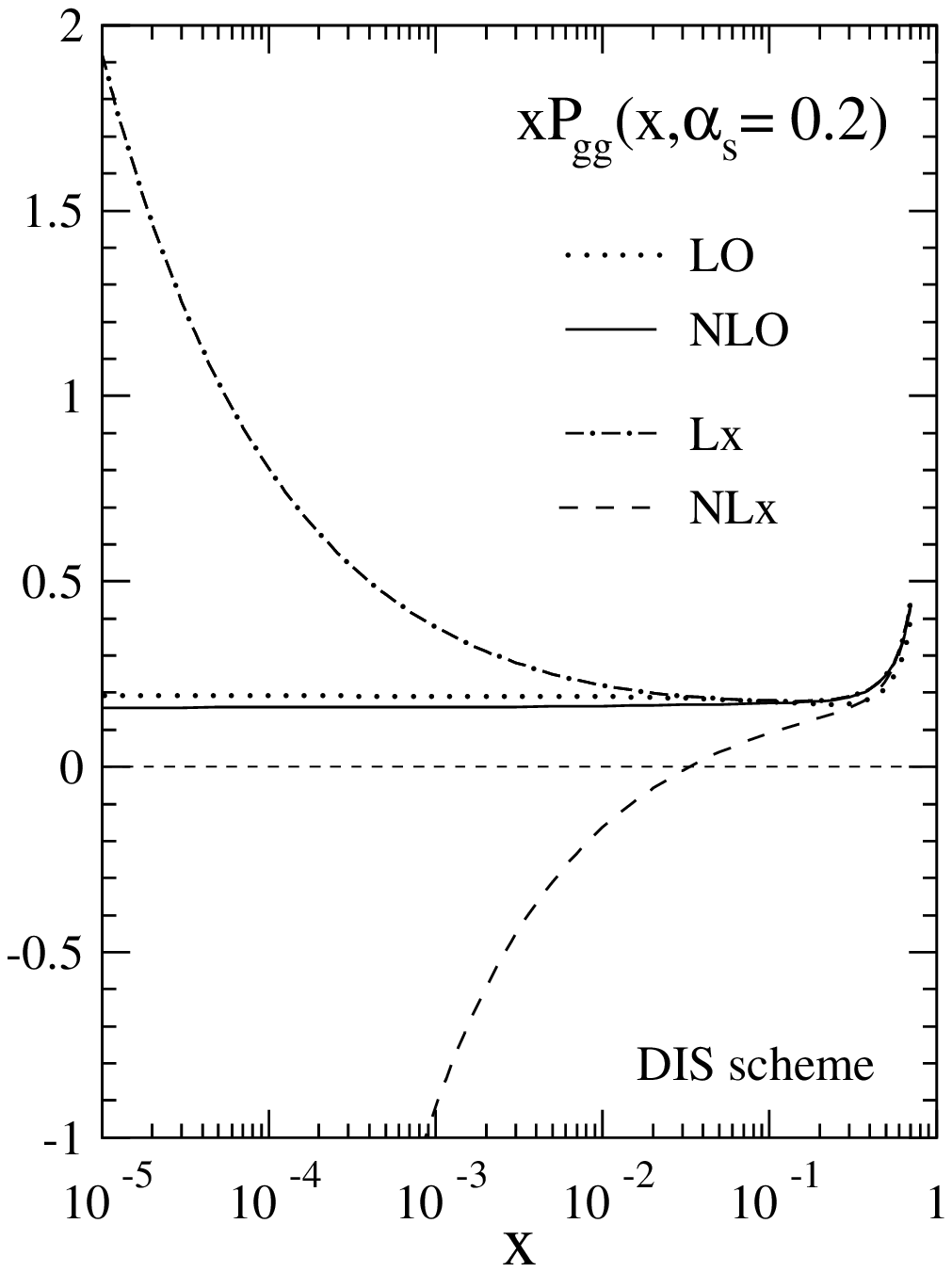,width=5.6cm,angle=0}}
}~~~
\parbox[c]{5.8cm}{
%\begin{table}[ht]
\begin{center}
\small
\vspace{-2mm}
\begin{tabular}{||r||r|r|r||}
\hline\hline
& & & \\[-3mm]
\multicolumn{1}{||c||}{$l$} &
\multicolumn{1}{c|}{$b^{\,\rm Lx}_{gg,l}$} &
\multicolumn{1}{c|}{$b^{\,\rm NLx}_{gg,l}$} &
\multicolumn{1}{c||}{ratio} \\
& & & \\[-3mm] \hline\hline
& & & \\[-3mm]
%----------------------------------------------------------------
 1 &  1.000$\,$E0 &  $-$1.139$\,$E0   &\hspace*{-1mm}  $-$1.1 \\
 2 &  0.000$\,$E0 &  $-$0.852$\,$E0   &\hspace*{-1mm}         \\
 3 &  0.000$\,$E0 &  $-$2.060$\,$E0   &\hspace*{-1mm}         \\
 4 &  2.404$\,$E0 &  $-$1.166$\,$E1   &\hspace*{-1mm}  $-$4.8 \\
 5 &  0.000$\,$E0 &  $-$1.844$\,$E1   &\hspace*{-1mm}         \\
 6 &  2.074$\,$E0 &  $-$4.416$\,$E1   &\hspace*{-1mm} $-$21.3 \\
 7 &  1.734$\,$E1 &  $-$1.712$\,$E2   &\hspace*{-1mm}  $-$9.9 \\
 8 &  2.017$\,$E0 &  $-$3.142$\,$E2   &\hspace*{-1mm}$-$155.8 \\
 9 &  3.989$\,$E1 &  $-$8.720$\,$E2   &\hspace*{-1mm} $-$21.9 \\
10 &  1.687$\,$E2 &  $-$2.835$\,$E3   &\hspace*{-1mm} $-$16.8 \\
11 &  6.999$\,$E1 &  $-$5.814$\,$E3   &\hspace*{-1mm} $-$83.1 \\
12 &  6.613$\,$E2 &  $-$1.706$\,$E4   &\hspace*{-1mm} $-$25.8 \\  
13 &  1.945$\,$E3 &  $-$5.004$\,$E4   &\hspace*{-1mm} $-$25.7 \\
14 &  1.718$\,$E3 &  $-$1.126$\,$E5   &\hspace*{-1mm} $-$65.5 \\
15 &  1.064$\,$E4 &  $-$3.335$\,$E5   &\hspace*{-1mm} $-$31.3 \\
%----------------------------------------------------------------
[0.5mm]\hline \hline
\end{tabular}
\normalsize
\end{center}
}
\\ \mbox{}
\end{minipage}

\noindent
{\small\sf Figure~2: The cumulative effect of the LO, NLO, L$x$ and 
NL$x$ contributions to the splitting function $P_{gg} (x,\alpha_s)$ in 
the DIS scheme at a typical value of $\alpha_s$. Also shown are the 
first fifteen L$x$ and NL$x$ $N$-space expansion coefficients 
$b_{gg,l}$ for four flavours \cite{BRNV}.} 
\vspace{-1.5mm}
\end{figure}

%%%%%%%%%%%%%%%%%%%%%%%%%%%%%%%%%%%%%%%%%%%%%%%%%%%%%%%%%%%%%%%%%%%%%%%
\section{Status of parton density parametrizations}
%%%%%%%%%%%%%%%%%%%%%%%%%%%%%%%%%%%%%%%%%%%%%%%%%%%%%%%%%%%%%%%%%%%%%%%

Major updates have been presented by the MRS~\cite{MRST} and GRV~\cite
{GRV} groups, superseding their respective '96 and '94 parton sets
\cite{old}. The CTEQ collaboration has released a dedicated study of 
the uncertainty of the gluon density~\cite{CTEQg}. The present 
(central) sets refer to values of $\als (M_z^2)$ = 0.116, 0.1175 and 
0.114 for the CTEQ$\,$4 \cite{CTEQ4}, MRST$\,$1 and GRV$\,$(98) 
distributions, respectively. Note also that heavy quarks are treated 
differently in these parametrizations. 

\vspace{1mm}
The present status of the gluon density is shown in Fig.~3. In the
small-$x$ part the preliminary H1 and ZEUS error bands from $F_2$
scaling violations~\cite{HERAg}, as well as the recent H1 results
from DIS charm production \cite{HERAc}, are compared to these
parametrizations. The difference between CTEQ4M (`massless charm') and 
CTEQ4F3~\cite{CTEQc} (`massive charm') indicates the impact of the 
heavy quark treatment. The small-$x$ gluon density seems rather well 
constrained down to $x \simeq 10^{-4}$. Note, however, that relevant
theoretical uncertainties (estimated, e.g., by factorization-scale 
variations) are not taken into account in Fig.~3.
\begin{figure}[h]
\begin{minipage}[b]{11.9cm}
\vspace*{-1.5mm}
\parbox[c]{5.8cm}{
\centerline{\epsfig{file=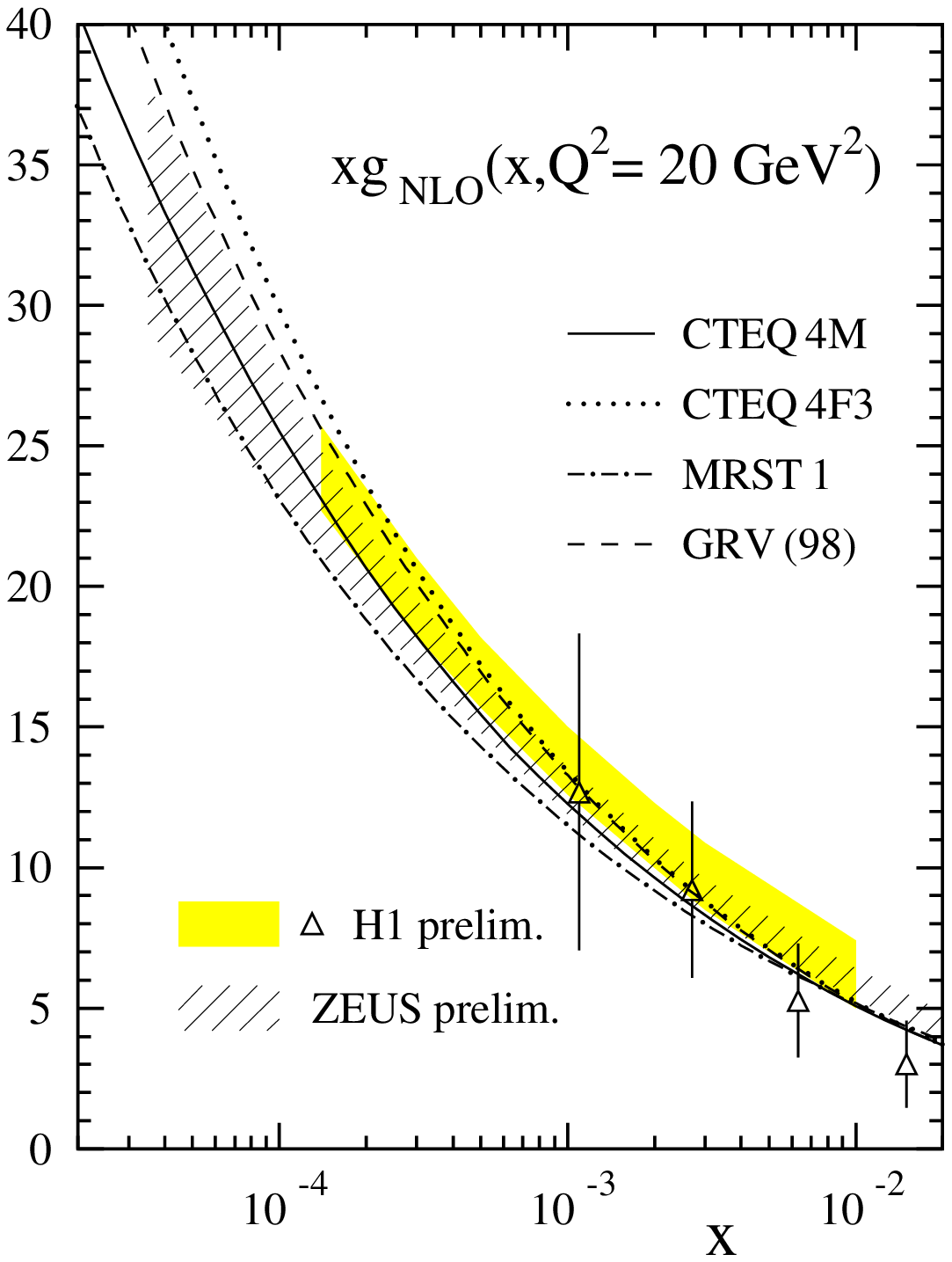,width=5.5cm,angle=0}}
}~~~
\parbox[c]{5.8cm}{
\centerline{\epsfig{file=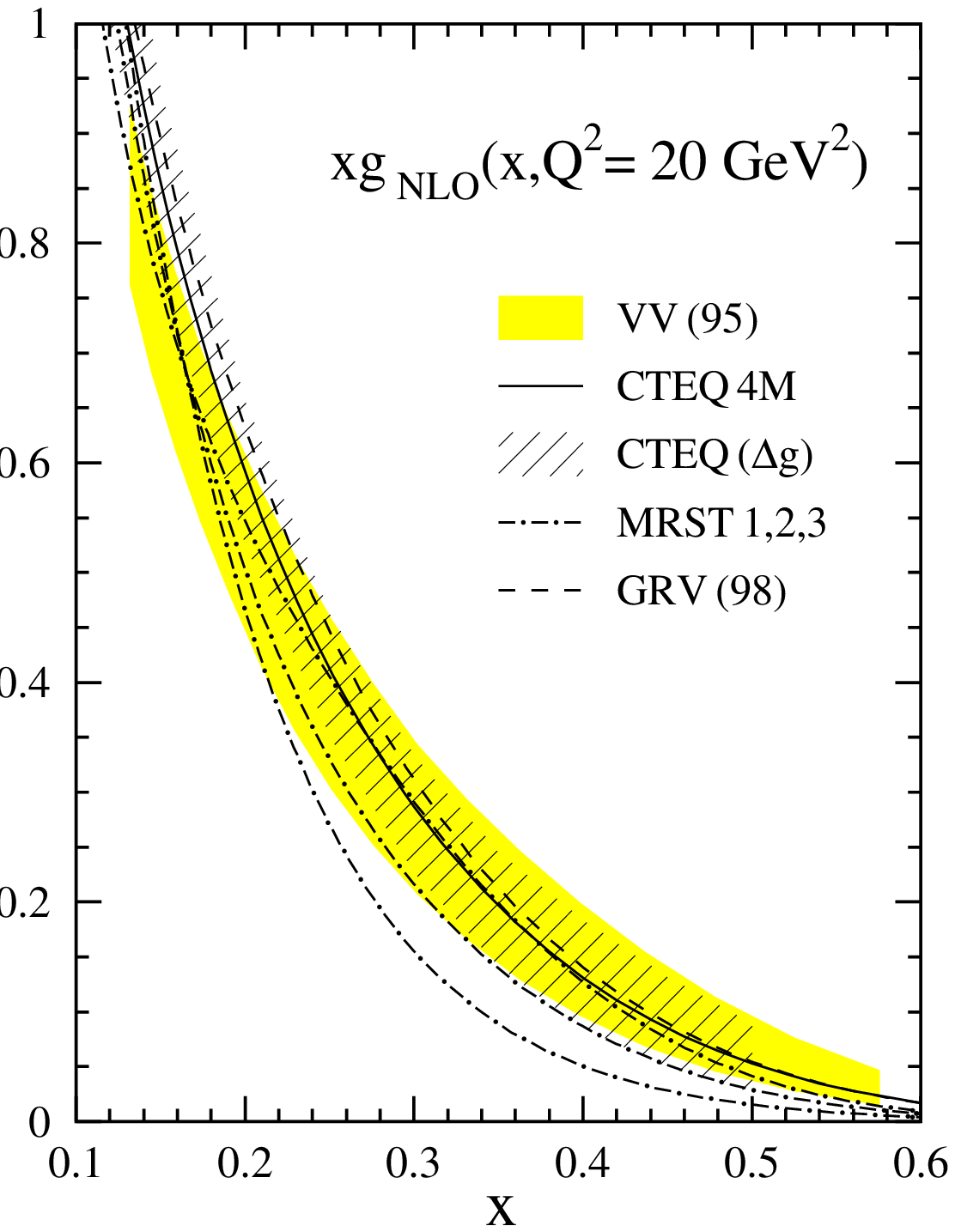,width=5.5cm,angle=0}}
}
\\ \mbox{}
\end{minipage}
 
\noindent
{\small\sf Figure~3: Present small-$x$ and large-$x$ constraints on the 
 proton's gluon distribution.}
\vspace{-1.0mm}
\end{figure}

Large uncertainties on $xg$ still persist in the large-$x$ region, 
$x \gsim 0.2$. Here the classic constraint has been prompt-photon 
production in $pp$ collisions~\cite{WA70}. However, extractions of the 
gluon density from these data suffer from sizeable scale uncertainties 
as shown by the gray error band~\cite{VV}. In addition, there is the 
possibility of a sizeable gluon $k_T$, as recently indicated by results 
from E706 \cite{E706}. In fact, the MRST large-$x$ error band~\cite
{MRST} on $xg$ stems from varying $k_T$ between 0 (upper curve, set 2) 
and 640 MeV (lowest curve, set 3) in the fit to the WA70 data. In view 
of these problems the CTEQ gluon-uncertainty analysis derives its error 
band from DIS and Drell-Yan data alone. It is interesting to note that 
all three bands are similarly wide for $x \gsim 0.3$. By propagation to 
high scales, benchmark uncertainties have been derived for gluon-gluon 
and gluon-quark luminosities at the {\sc Tevatron} and the LHC
\cite{CTEQg}.
 
\vspace{1mm}
As the total quark density is quite well constrained by $F_2$ data for 
$10^{-4} \lsim x $ $\lsim 0.7$, the other critical issues are the
flavour decomposition and the $x \ra 1$ behaviour. In both areas new
results have been reported. E866 has published their high-mass data on
the $pp$/$pd$ Drell-Yan asymmetry~\cite{E866}. As shown in Fig.~4, 
these results strongly constrain the ratio $\bar{d}/\bar{u}$, 
especially in the range $0.1 \lsim x \lsim 0.3$. Preliminary results on 
this ratio, inferred from semi-inclusive DIS, as well as data on 
$F_2^n / F_2^p$ have also been presented by HERMES~\cite{Ouy}.
\begin{figure}[h]
\begin{minipage}[b]{11.9cm}
\vspace{-1mm}
\parbox[c]{5.8cm}{
\centerline{\epsfig{file=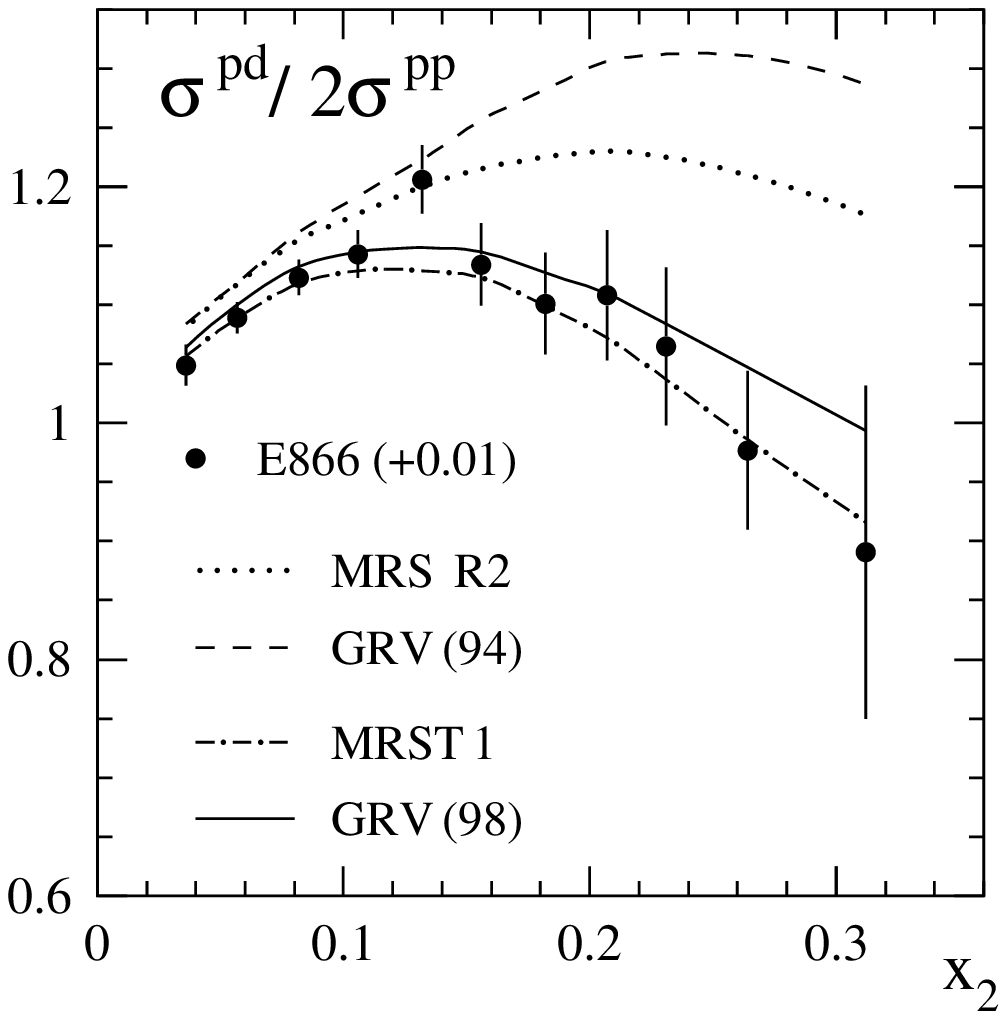,width=5.7cm,angle=0}}
}~~~
\parbox[c]{5.8cm}{
\centerline{\epsfig{file=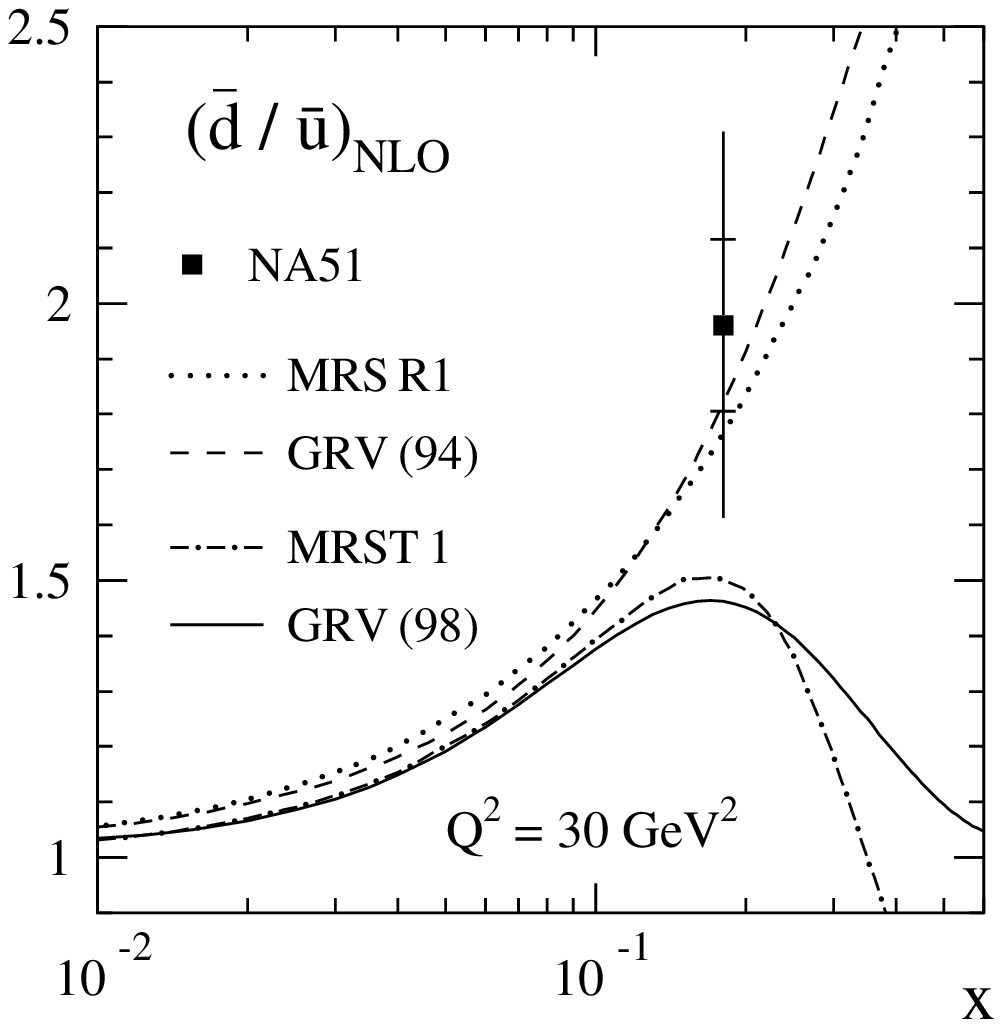,width=5.7cm,angle=0}}
}
\\ \mbox{}
\end{minipage}
 
\noindent
{\small\sf Figure~4: The E866 Drell-Yan asymmetry data and their impact
 on the light-quark sea.}
\vspace{-0.5mm}
\end{figure}

$F_2^n / F_2^p$ is a dominant source of information on the $d/u$ ratio. 
As $F_2^n$ is inferred from deuteron measurements, possible nuclear 
binding effects enter here~\cite{FSMT}. Such effects have been 
neglected in recent parton parametrization. They can, however, lead to
drastic modifications of the valence-quark ratio $d_v / u_v$ at large
$x$, as quantified in ref.~\cite{YB}. The region of very large $x$, 
$x \gsim 0.8$, is usually not taken into account either. The DIS and 
resonance-region data of SLAC have been employed to study the absolute 
$x \ra 1$ behaviour of the valence quarks. A significant flat  
contribution, as previously discussed in connection with the HERA high 
$Q^2$ events~\cite{KLT}, is found to be strongly disfavoured~\cite{YB}.

%%%%%%%%%%%%%%%%%%%%%%%%%%%%%%%%%%%%%%%%%%%%%%%%%%%%%%%%%%%%%%%%%%%%%%%
\section{Higher-twist effects and the strong coupling constant}
%%%%%%%%%%%%%%%%%%%%%%%%%%%%%%%%%%%%%%%%%%%%%%%%%%%%%%%%%%%%%%%%%%%%%%%

Present-day determinations of $\als $ from DIS structure functions
involve relatively low values of $Q^2$. Hence higher-twist corrections 
can have appreciable effects. A new result on $\als(M_z^2)$ from the 
Gross-LLewellyn-Smith (GLS) sum rule has been reported by CCFR 
\cite{Yu}. Their iron data, together with results from other neutrino
experiments, in the region $ 1.3 \leq Q^2/\mbox{GeV}^2 \leq 5.0$ lead 
to
%$\,$\footnote
%{Note that some $\als $-values given here have been significantly 
%changed wrt.\ the presentations at the workshop.} 
$$
  \als^{\rm NNLO}(M_z^2) _{\rm GLS} = 0.114^{\:\: +0.009}_{\:\: -0.011} 
  \mbox{ (exp.)} \pm 0.05 \mbox{ (th.)} \: . 
$$ 
The theoretical error is dominated by the uncertainty of the 
higher-twist contribution. Nuclear $1/Q^2$ corrections have been 
studied for this case in ref.~\cite{Kul}. Very small corrections to the
GLS sum rule are found, but sizeable effects for the incomplete GLS
integral, $S_{\rm GLS}(x \!\neq\! 0,Q^2)$, and for the iron-to-nucleon
ratio $R_3(x,Q^2) = F_3^{\,\rm Fe} /F_3^{\,\rm N}$, see Fig.~5. The 
latter results may be relevant for $\als$ determinations from scaling 
violations in neutrino-nucleus DIS \cite{CCFRa}.
\begin{figure}[h]
\begin{minipage}[b]{11.9cm}
\vspace{-1mm}
\parbox[c]{5.8cm}{
\centerline{\epsfig{file=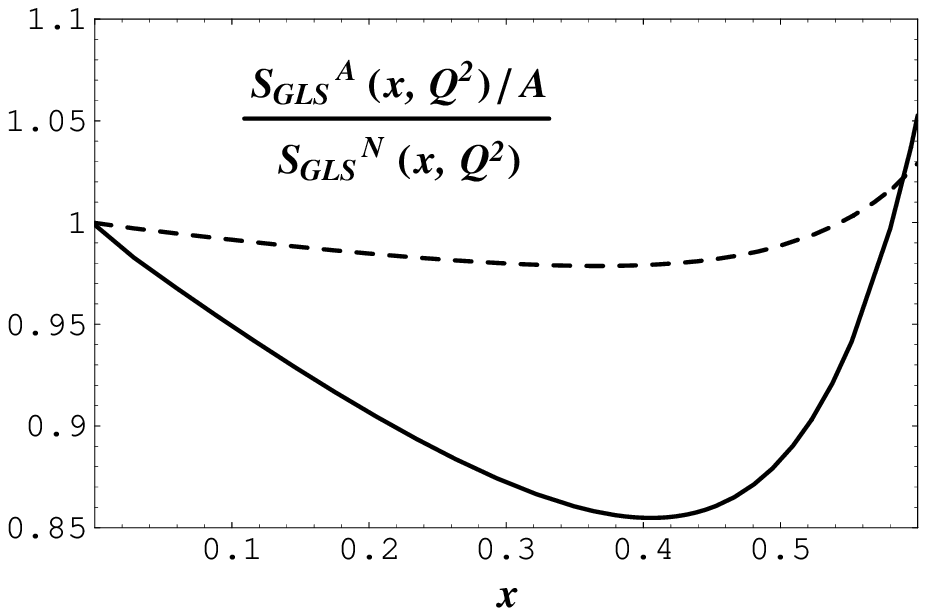,width=5.8cm,angle=0}}
}~~~
\parbox[c]{5.8cm}{
\centerline{\epsfig{file=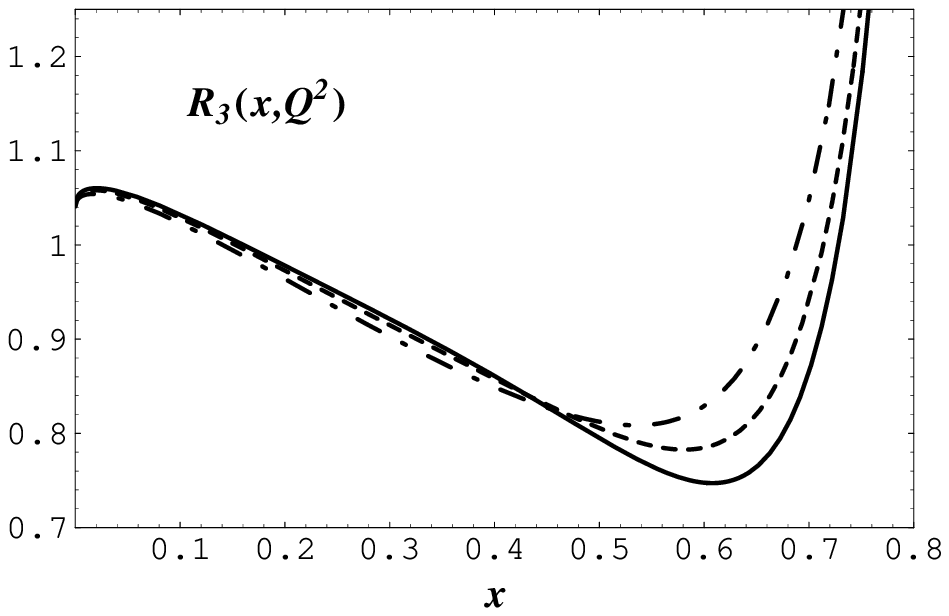,width=5.8cm,angle=0}}
}
\\ \mbox{}
\end{minipage}
 
\noindent
{\small\sf Figure~5: Predictions for the nuclear effects in $F_3
 ^{\,\nu A}$~\cite{Kul}. Left part: the incomplete GLS integral at 
 $Q^2 = 5 \mbox{ GeV}^2$ for A = d (dashed) and A = Fe (solid). 
 Right part: the ratio $R_3$ for iron at $Q^2$ = 3 (solid), 5 (dashed), 
 and 15 GeV$^2$ (dashed-dotted).}
\vspace{-0.5mm}
\end{figure}

New fits to the electromagnetic DIS data of BCDMS, SLAC and NMC at 
$x > 0.3$ have been presented as well~\cite{KK}, including target-mass 
corrections and simple approximations for dynamical higher-twist terms. 
If the shape of the latter is allowed to vary freely, the fits result 
in 
$$
  \als^{\rm NLO}(M_z^2)_{\partial_Q F_2^{\,\rm e.m.}} = 0.114 \pm 0.002 
  \mbox{ (exp.)} \: .
$$
Not surprisingly, this result agrees with previous analyses~\cite{MV}.
On the other hand, a very low value of $\als (M_z^2) = 0.103 \pm 0.002
\mbox{ (exp.)}$ is obtained~\cite{KK} if the shapes of the twist-4 and 
twist-6 terms are adopted from the renormalon model~\cite{DW}. Note 
that this finding seems to be at variance with the analysis of ref.\ 
\cite{YB}, where good agreement between the large-$x$ data and the 
renormalon approach has been found for a fixed value of $\als(M_z^2) 
= 0.120$.

\section*{Acknowledgments}
It is a pleasure to thank Pierre Marage and Robert Roosen for their
help and kind hospitality in Brussels. This work was supported by the 
German Federal Ministry for Research and Technology under contract 
No.~05~7WZ91P(0).

%%%%%%%%%%%%%%%%%%%%%%%%%%%%%%%%%%%%%%%%%%%%%%%%%%%%%%%%%%%%%%%%%%%%%%%


\section*{References}
\begin{thebibliography}{99}
%
%[1]
\bibitem{WG1a} L. Demortier, A. Mehta, and W. Zeuner, these 
proceedings.
%-----------------------------------------------------------------------
%
%[2]
\bibitem{LSRN}
E.\ Laenen, S.\ Riemersma, J.\ Smith and W.L.\ van Neerven, {\em Nucl.\ 
Phys.} {\bf B392} (1993) 162.
%-----------------------------------------------------------------------
%
%[3]
\bibitem{ASYM}
M. Buza, Y. Matiounine, J. Smith, R. Migneron, and W.L. van Neerven,
{\em Nucl.\ Phys.} {\bf B472} (1996) 611.
%-----------------------------------------------------------------------
%
%[4]
\bibitem{ACOT} M. Aivazis, F. Olness, and W.-K. Tung, {\em Phys.\ Rev.}
{\bf D50} (1994) 3085; M. Aivazis, J. Collins, F. Olness, and 
W.-K. Tung, {\em ibid.} 3102. 
%-----------------------------------------------------------------------
%
%[5]
\bibitem{F4N}
M. Buza, Y. Matiounine, J. Smith, and W.L. van Neerven, {\em Eur.\ 
Phys.\ J.} {\bf C1} (1998) 301; {\em Phys.\ Lett.} {\bf B411} (1997) 
211; W.L. van Neerven, these proceedings ({\tt hep-ph/9804445}). 
%-----------------------------------------------------------------------
%
%[6]
\bibitem{RT} R.G. Roberts and R.S. Thorne, {\em Phys.\ Rev.} {\bf D57}
(1998) 6871; {\em Phys.\ Lett.} {\bf B421} (1998) 303; R.S. Thorne, 
these proceedings ({\tt hep-ph/9805298}). 
%-----------------------------------------------------------------------
%
%[7]
\bibitem{OS}
F.I. Olness and R.J. Scalise, {\em Phys.\ Rev.} {\bf D57} (1998) 241.
%-----------------------------------------------------------------------
%
%[8]
\bibitem{AVc} A. Vogt, Proceedings of DIS '96, Rome, April 1996, eds.\ 
G. D'Agostini and A.\ Nigro (World Scientific 1997), p. 254 
({\tt hep-ph/9601352}).
%-----------------------------------------------------------------------
%
%[9]
\bibitem{HS} B.W. Harris and J. Smith, {\em Nucl.\ Phys.} {\bf B452} 
(1995) 109; {\em Phys.\ Rev.} {\bf D57} (1998) 2806.
%-----------------------------------------------------------------------
%
%[10]
\bibitem{Ol1} J. Amundson, F. Olness, C. Schmidt, W.-K. Tung, and 
 X. Wang, these proceedings.
%-----------------------------------------------------------------------
%
%[10]
\bibitem{Moch} S. Moch, these proceedings ({\tt hep-ph/9805370}).
%-----------------------------------------------------------------------
%
%[11]
\bibitem{LNRV}
S.A. Larin, P. Nogueira, T. van Ritbergen, and J.A.M. Vermaseren,
{\em Nucl.\ Phys.} {\bf B492} (1997) 338.
%-----------------------------------------------------------------------
%
%[12]
\bibitem{MSN}
Y. Matiounine, J. Smith, W. van Neerven, {\em Phys.\ Rev.} {\bf D57} 
(1998) 6701.
%-----------------------------------------------------------------------
%
%[13]
\bibitem{BG}
J.F. Bennett and J.A. Gracey, {\em Nucl.\ Phys.} {\bf B517} (1998) 241;
J.A.~Gra\-cey, these proceedings ({\tt hep-ph/9805310}).
%-----------------------------------------------------------------------
%
%[14]
\bibitem{CH}
S. Catani and F. Hautmann, {\em Nucl.\ Phys.} {\bf B427} (1994) 475.
%-----------------------------------------------------------------------
%
%[15]
\bibitem{EHW}
R.K. Ellis, F. Hautmann, and B. Webber, {\em Phys.\ Lett.} {\bf B348}
(1995) 582;\\
J. Bl\"umlein, S. Riemersma,  and A. Vogt, {\em Nucl.\ Phys.} {\bf B}
(Proc. Suppl.) {\bf 51C} (1996) 30.
%-----------------------------------------------------------------------
%
%[16]
\bibitem{Th2}
R.S. Thorne, {\em Phys.\ Lett.} {\bf B392} (1997) 463; {\em Nucl.\ 
Phys.} {\bf B512} (1998) 323; these proceedings ({\tt hep-ph/9805299}).
%-----------------------------------------------------------------------
%
%[17]
\bibitem{Mun}
S. Munier, R. Peschanski, {\em Nucl.\ Phys.} {\bf B524} (1998) 377; 
S. Munier, these proceedings ({\tt hep-ph/9805299}).
%-----------------------------------------------------------------------
%
%[18]
\bibitem{FL}
V.S. Fadin and L.N. Lipatov, {\tt hep-ph/9802290}; V.S. Fadin, these
proceedings.
%-----------------------------------------------------------------------
%
%[19]
\bibitem{CC}
G. Camici and M. Ciafaloni, {\em Nucl.\ Phys.} {\bf B496} (1997) 305; 
{\em Phys.\ Lett.} {\bf B412} (1997) 396 (E: {\bf B417}(1998) 390); 
{\tt hep-ph/9803389}; M. Ciafaloni, these proceedings.
%-----------------------------------------------------------------------
%
%[20]
\bibitem{BRNV}
J. Bl\"umlein and A. Vogt, {\em Phys.\ Rev.} {\bf D57} (1998) 1;
{\em Phys.\ Rev.} {\bf D58} (1998) 014020; J. Bl\"umlein, V. Ravindran, 
W.L. van Neerven, and A.~Vogt, these proceedings ({\tt hep-ph/9806368}).
%-----------------------------------------------------------------------
%
%[21]
\bibitem{BF}
R.D. Ball and S. Forte, these proceedings ({\tt hep-ph/9805315}).
%-----------------------------------------------------------------------
%
%[22]
\bibitem{MRST}
A.D. Martin, R.G. Roberts, W.J. Stirling, and R.S. Thorne, 
{\em Eur.\ Phys.\ J.} {\bf C4} (1998) 463; these proceedings
({\tt hep-ph/9805205}).
%-----------------------------------------------------------------------
%
%[23]
\bibitem{GRV}
M. Gl\"uck, E. Reya, and A. Vogt, {\tt hep-ph/9806404}; A. Vogt, talk 
presented at this workshop.
%-----------------------------------------------------------------------
%
%[24]
\bibitem{old}
A.D. Martin, R.G. Roberts, W.J. Stirling, {\em Phys.\ Lett.} 
{\bf B387} (1996) 419; \\ 
M. Gl\"uck, E. Reya, and A. Vogt, {\em Z. Phys.} {\bf C67} (1995) 433. 
%-----------------------------------------------------------------------
%
%[25]
\bibitem{CTEQg}
J. Huston et al., CTEQ Coll., {\tt hep-ph/9801444}; these proceedings.
%-----------------------------------------------------------------------
%
%[26]
\bibitem{CTEQ4}
H.L. Lai et al., CTEQ Coll., {\em Phys.\ Rev.} {\bf D55} (1997) 1280.
%-----------------------------------------------------------------------
%
%[27]
\bibitem{HERAg}
H1 Coll., EPS-Conference Jerusalem, August 1997, paper 260.; \\
ZEUS Coll., ibid., paper N-647.
%-----------------------------------------------------------------------
%
%[28]
\bibitem{HERAc}
H1 Coll., EPS-Conference Jerusalem, August 1997, paper 275. 
%-----------------------------------------------------------------------
%
%[29]
\bibitem{CTEQc}
H.L. Lai and W.K. Tung, {\em Z. Phys.} {\bf C74} (1997) 463.
%-----------------------------------------------------------------------
%
%[30]
\bibitem{WA70}
W. Vogelsang and M.R.\ Whalley, {\em J.\ Phys.} {\bf G23}, Suppl.\ 7A 
(1997) A1.
%M. Bonesini et al., WA70 Coll., {\em Z. Phys.} {\bf C38} (1988) 371.
%-----------------------------------------------------------------------
%
%[31]
\bibitem{VV}
W. Vogelsang and A. Vogt, {\em Nucl.\ Phys.} {\bf B453} (1995) 334. 
%-----------------------------------------------------------------------
%
%[32]
\bibitem{E706}
L. Apanasevich et al., E706 Coll., {\tt hep-ex/9711017}.
%-----------------------------------------------------------------------
%
%[33]
\bibitem{E866}
E.A. Hawker et al., E866 Coll., {\em Phys.\ Rev.\ Lett.} {\bf 80} 
(1998) 3715;  \\ L.D. Isenhower, these proceedings.
%-----------------------------------------------------------------------
%
%[34]
\bibitem{Ouy}
J. Ouyang, HERMES Coll., these proceedings.
%-----------------------------------------------------------------------
%
%[35]
\bibitem{FSMT}
L.L. Frankfurt and M.I. Strikman, {\em Phys.\ Rept.} {\bf 160} (1988) 
235; \\
W. Melnitchouk and A.W. Thomas, {\em Nucl.\ Phys.} {\bf A631} (1998) 
296c.
%-----------------------------------------------------------------------
%
%[36]
\bibitem{YB}
U.K. Yang and A. Bodek, these proceedings ({\tt hep-ph/9806458}).
%-----------------------------------------------------------------------
%
%[37]
\bibitem{KLT}
S. Kuhlmann, H.L. Lai, and W.-K. Tung, {\em Phys.\ Lett.} {\bf B409} 
(1997), 271.
%-----------------------------------------------------------------------
%
%[38]
\bibitem{Yu}
J. Yu, CCFR/NuTeV Coll., these proceedings ({\tt hep-ex/9806031}).
%-----------------------------------------------------------------------
%
%[39]
\bibitem{Kul}
S.A. Kulagin, {\tt nucl-th/9801039}; these proceedings.
%-----------------------------------------------------------------------
%
%[40]
\bibitem{CCFRa}
W.G. Seligman et al., CCFR Coll., {\em Phys.\ Rev.\ Lett.} {\bf 79}
(1997) 1213.
%-----------------------------------------------------------------------
%
%[41]
\bibitem{KK}
A. Kotikov and V. Krivokhijine, these proceedings 
({\tt hep-ph/9805353}). 
%-----------------------------------------------------------------------
%
%[42]
\bibitem{MV}
A. Milsztajn and M. Virchaux, {\em Phys.\ Lett.} {\bf B274} (1992) 221.
%-----------------------------------------------------------------------
%
%[43]
\bibitem{DW}
M. Dasgupta and B.R. Webber {\em Phys.\ Lett.} {\bf B382} (1996) 273.
%-----------------------------------------------------------------------
\end{thebibliography}
\end{document}